\documentclass[prb,aps,twocolumn,superscriptaddress]{revtex4-1}
\usepackage{graphicx}
\usepackage{dcolumn}
\usepackage{bm}
\usepackage{bbm}
\usepackage{amsfonts}
\usepackage{amsmath}
\usepackage{natbib}
\providecommand{\keywords}[1]{\textbf{\textit{Index terms---}} #1}
\usepackage[mathlines]{lineno}
\usepackage[colorinlistoftodos]{todonotes}
\usepackage[colorlinks=true, allcolors=blue]{hyperref}

\begin{document}

\newcommand{\unamone}{Departamento de Sistemas Complejos, Instituto de Fisica,
Universidad Nacional Aut\'onoma de M\'exico, Apartado Postal 20-364,01000,
Ciudad de M\'exico, M\'exico.}
\newcommand{\unamtwo}{Instituto de F\'isica,
Universidad Nacional Aut\'onoma de M\'exico, Apartado Postal 20-364 01000,
Ciudad de  M\'{e}xico, M\'{e}xico}
\newcommand{\uama}{\'Area de F\'isica Te\'orica y Materia Condensada,
Universidad Aut\'onoma Metropolitana Azcapotzalco, Av. San Pablo 180,
Col. Reynosa-Tamaulipas, 02200 Cuidad de M\'exico, M\'exico}

\title{
Dynamical band gap tuning in Weyl semi-metals by intense elliptically
polarized normal illumination and its application to $8-Pmmn$ borophene.
}
\author{V. G. Ibarra-Sierra}
\affiliation{\unamone}
\author{J. C. Sandoval-Santana}
\affiliation{\unamtwo}
\author{A. Kunold}
\affiliation{\uama}
\author{Gerardo G. Naumis}
\affiliation{\unamone}

\date{\today}

\begin{abstract}

The Dynamical-gap formation in
Weyl semimetals
modulated by intense elliptically polarized light 
is addressed through the solution
of the time-dependent Schr\"odinger equation
for the Weyl Hamiltonian
via the Floquet theorem.
The time-dependent wave functions and the quasi-energy spectrum
of the two-dimensional Weyl Hamiltonian under
normal incidence of elliptically polarized electromagnetic   
waves are obtained using a non-perturbative approach.
In it, the Weyl equation is reduced to an ordinary
second-order differential Mathieu equation.
It is shown that the stability conditions of the Mathieu functions
are directly inherited by the wave function
resulting in a quasiparticle spectrum
consisting of bands and gaps determined by dynamical diffraction and 
resonance conditions between the electron and the electromagnetic wave.
Estimations of the electromagnetic field intensity and frequency,
as well as the magnitude of the generated gap are obtained
for the $8-Pmmn$ phase of borophene.
We provide with a simple method that enables to predict the formation
of dynamical-gaps of unstable wave functions and their magnitudes.
This method can readily be adapted to other Weyl semimetals.

\begin{description}
\item[Keywords]
Weyl semi-metals; dynamical gap; Mathieu equation 
\end{description}
\end{abstract}

\maketitle

\section{Introduction}

Recently, the so-called
Dirac and Weyl materials
have received considerable attention due to their
possible implementation into next-generation electronic devices 
\cite{novoselov2012roadmap,peng2016electronic,ferrari2015science,
naumis2017electronic}.
The main physical properties of these materials were observed for the 
first time in graphene, an allotrope of carbon 
consisting of a monolayer of atoms in a honeycomb lattice with an 
electron linear dispersion near the Dirac points. As a result
of this characteristics,
the charge carriers in graphene behave like massless Dirac fermions 
\cite{geim2009graphene,neto2009electronic,oliva2014anisotropic,
naumis2014mapping,oliva2015generalizing,oliva2015tunable,
oliva2015sound,naumis2017electronic,carrillo2018band}.
Thereafter, a wide variety of two-dimensional materials with similar
properties has been discovered \cite{wehling2014dirac}.
Examples of these are: silicene
\cite{jose2013structures, liu2013silicene},
germanene \cite{behera2011first,zhang2016structural},
stanene \cite{zhu2015epitaxial,chen2016electronic} and
artificial graphene \cite{soltan2011multi,gomes2012designer}.
Borophen, a two dimensional allotrope of boron, also
falls in this category.
The chemical similarity between boron and carbon atoms
has triggered the search for stable two-dimensional
boron structures and synthesis techniques to
produce them\cite{zhang2017two}.
Since their theoretical prediction \cite{BOUSTANI1997355},
many different allotropes of borophene have been
experimentally confirmed \cite{Mannix1513}.
Among its many different phases,
the orthorhombic $8-Pmmn$ is one of the most energetically stable
structures \cite{peng2016electronic},
having a ground state energy lower than its analogs 
\cite{verma2017effect}.
Borophene, in contrast with graphene, shows a highly anisotropic 
crystalline structure, which causes high optical anisotropy
and transparency
\cite{peng2016electronic,verma2017effect,Champo2019}.
It is thus a strong candidate for flexible
electronics, display technologies and in the design
of smart windows where minimal photon absorption and reflection
are required
\cite{zhang2016correlated, peng2016electronic,zhang2017two}.

Despite the many useful and fascinating properties
of graphene, borophene and Weyl materials in general,
their lack of an electronic 
band gap has stimulated the search for either other
two-dimensional materials with
semiconducting properties or techniques to induce them artificially.
Among other proposals to circumvent this problem,
one of the most promising ideas is generating a light
induced dynamical-gap. As the electromagnetic field
is a periodical function of time, this technique has been
termed Floquet gap engineering.
High intensity electromagnetic waves interacting with
graphene have  been studied using perturbative approaches 
\cite{lopez2010graphene,higuchi2017light}. However, it has
been shown that light induces a renormalization of the electronic
spectrum of Dirac materials not captured by simple perturbation
techniques \cite{lopez2008analytic,lopez2010graphene,
kibis2010metal,kristinsson2016control,oliva2016effective,
kibis2015,kibis2017all}.
In this regard, borophene brigns interesting possibilities
to study the light-matter interaction in Weyl semimetals due
to its asymmetric spectrum.
As graphene, borophene has a honeycomb lattice with
two nonequivalent sublattices. However, its
peculiar structure give rise to a tilted anisotropic cone
in the vicinity of the Dirac points
\cite{Islam2017,verma2017effect,Champo2019}, as opposed to
graphene whose spectrum is completely isotropic in $K$ space.

Recently, the formation of energy gaps in borophene subject
high-intensity linearly
polarized light was studied beyond the
perturbative approximation \cite{Champo2019}.
It was found that borophene, when interacting
with light, acquires a complex band structure
from the stability conditions of the solutions of the Mathieu
differential equation.
Among other effects, the interaction with light
produces a gap in the vicinity of borophene's Dirac point.
The effects of an intense circularly polarized
electromagnetic field have, nevertheless, not been discussed
for borophene yet.

\begin{figure}[t]
\begin{center}
\includegraphics[scale=0.65]{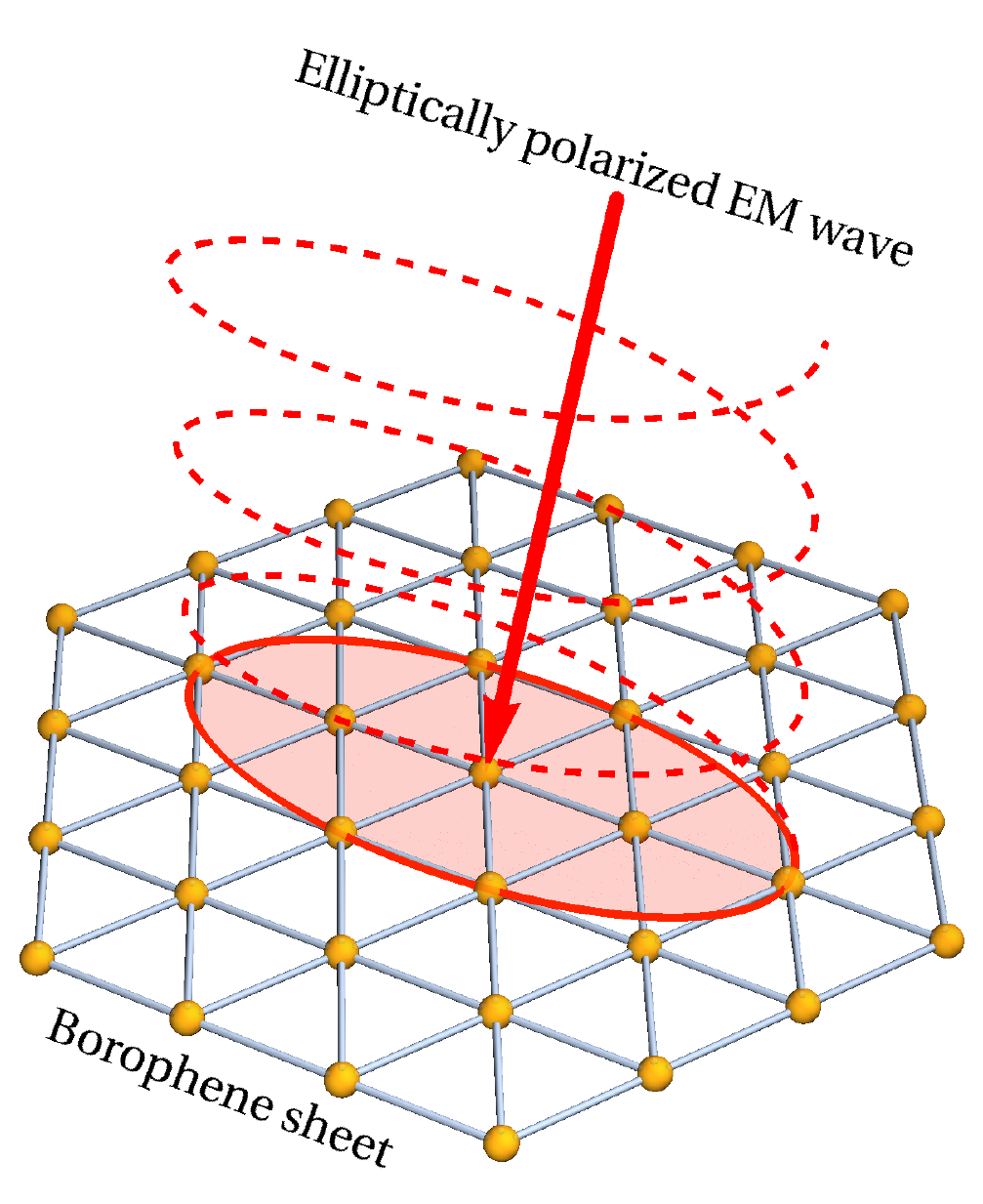}
\end{center}
\caption{\label{Fig:BoropheneLattice} Borophene sheet under an elliptically polarized electromagnetic wave.}
\end{figure}

In this paper, we address the general problem of a particle that obeys the
Weyl Hamiltonian subject to an intense elliptically polarized 
electromagnetic field.
As it is schematized in Fig. \ref{Fig:BoropheneLattice},
our solutions can be applied to the particular case of electrons in
borophene under a strong elliptically polarized field. 
We report the wave functions, the quasi-energy spectrum,
and the magnitude of the dynamic gap opening.
The case of linearly polarized light,
addressed by us previously \cite{Champo2019},
is proven to be fundamentally different from the elliptically
polarized one studied in this work.
Our analysis mainly focuses on the stability
and instability of the time-dependent wave functions.  
The results presented here display an interesting interplay between the tilted
anisotropy and the relative orientation of the
light-polarization ellipse.
Moreover, we show that the gaps
may be tuned by changing
the orientation of the elliptical polarization profile of light.

The paper is organized as follows. In Sec. \ref{sec:DiracHamiltonian} we 
introduce the low-energy effective two-dimensional Weyl Hamiltonian
under an arbitrary electromagnetic field.
Subsequently, in Sec.  \ref{sec:BoropheEllipticallyPolarized}, we
determine the time-dependent wavefunction of electrons in borophene subject to an 
elliptically polarized electromagnetic field.
In this same section we analyze the stability of the solutions inherited from
Mathieu functions in the strong electromagnetic field or long wavelength 
regimes.
We workout the time-dependent wave functions and the solutions' stability 
chart. To get an insight into the gap structure,
in  Sec.  \ref{sec:quasienergy},
the stability and instability regions
are projected onto the tilted Dirac cones
of the free Weyl electrons.
In Sec. \ref{sec:quasienergy},
we extract the quasi-energy spectrum from the
time-dependent wave function and prove that it
consistently shows a similar gap structure to
that of the projected chart.
Finally, we summarize and conclude in Sec. \ref{sec:Conclusions}.

\section{Weyl electrons subject to electromagnetic
fields}\label{sec:DiracHamiltonian}

\subsection{The Weyl Hamiltonian}
The single-particle low-energy effective Weyl Hamiltonian
is given by 
\cite{Zabolotskiy2016,Islam2017,verma2017effect,Champo2019},
\begin{equation}\label{ec:AnsotropicDiracHamiltonian}
\hat{H}=\mu\left(v_t\hat{P}_y\hat{\sigma}_0
  +v_x\hat{P}_x\hat{\sigma}_x +v_y\hat {P}_y\hat{\sigma}_y\right),  
\end{equation}
where $\hat{P}_{x}$ and $\hat{P}_{y}$ are the momentum operators, 
$\hat{\sigma}_i$ are the  Pauli matrices,
and $\hat{\sigma}_0$ is the  $2\times 2$ identity matrix.
For $8-Pmmn$ borophene,
the three velocities in the Weyl
Hamiltonian (\ref{ec:AnsotropicDiracHamiltonian})
are $v_x=0.86v_F$, $v_y=0.69v_F$ and $v_t=0.32v_F$ in
units of the Fermi velocity $v_F=10^6\,m/s$ \cite{Zabolotskiy2016}.
The two Dirac points $\mathbf{k}=\pm\mathbf{k}_D$ are given
by the valley index $\mu=\pm 1$. The first term in
Eq. (\ref{ec:AnsotropicDiracHamiltonian}) gives rise to the tilting
of the Dirac cones and the last ones correspond
to the kinetic energy.

The previous Hamiltonian results in the energy dispersion relation
\cite{Islam2017}
\begin{equation} \label{ec:EnergyDispersion}
E_{\eta,k}^{\mu} = \mu \hbar v_{t} k_{y}+\eta\epsilon,
\end{equation}
where 
\begin{equation}\label{ec:EpsilonCoefficient}
\epsilon=\hbar\sqrt{v_{x}^{2} k_{x}^{2}+v_{y}^{2}k_{y}^{2}}.    
\end{equation}
The corresponding free Weyl electron wave function is,
\begin{equation}\label{ec:WaveFunctionFree}
\psi_{\eta,\,\bf{k}}^{\mu}(\mathbf{r})
  = \mu \frac{\exp(i\bf{k\cdot r})}{\sqrt{2}}
  \left[
  \begin{array}{lcc}
     1 \\
     \eta \exp(i \Theta) 
    \end{array} \right],
\end{equation}
where $\eta= \pm 1$ is the band index,
$\Theta= \tan^{-1} (v_{y}k_{y}/v_{x}k_{x})$ and the two-dimensional
momentum vector is given by $\mathbf{k}=(k_{x},k_{y})$. 

\subsection{The Weyl Hamiltonian in the presence of
an electromagnetic wave}

Now we consider a charge carrier described by the Weyl Hamiltonian
subject to an electromagnetic wave that propagates along a direction
perpendicular to the surface of the crystal.
From Eq. (\ref{ec:AnsotropicDiracHamiltonian}) and using the
minimal coupling we obtain, 
\begin{equation}\label{ec:DiracHamiltonianUnderElectromagneticField} 
\hat{H}= \left( 
\begin{array}{lcc}
v_{t} \hat{\Pi}_{y} & v_{x} \hat{\Pi}_{x}- i v_{y} \hat{\Pi}_{y} \\
v_{x} \hat{\Pi}_{x}+ i v_{y} \hat{\Pi}_{y} & v_{t} \hat{\Pi}_{y}
\end{array} 
\right),
\end{equation}
where $\hat{\boldsymbol{\Pi}}=\hat{\boldsymbol{P}}-e{\boldsymbol{A}}$, with
$\boldsymbol{A}=(A_x,A_y)$
being the vector potential of the incident electromagnetic 
wave. Calculations are considerably simplified by choosing
a gauge in which the vector potential is only a function of time.
The Schr\"odinger equation for charge carriers in a Weyl semimetal
is thus given by
\begin{equation}\label{ec:DiracEquationOne}
\hat{H}(\mathbf{r},t){\boldsymbol{\Psi}}(\boldsymbol{r},t)
= i \hbar \frac{\partial}{\partial t} {\boldsymbol{\Psi}}(\boldsymbol{r},t),
\end{equation}
where, in the two dimensional spinor
$\boldsymbol{\Psi}(\mathbf{r},t)
 =\left(\Psi_{A}(\mathbf{r},t),\Psi_{B}(\mathbf{r},t)\right)^{\top}$,
$A$ and $B$ label the two sublattices.

To deduce the explicit form of the wave function ${\bf{\Psi}}(\mathbf{r},t)$
from Eq. (\ref{ec:DiracEquationOne})
we make the following ansatz
\begin{equation} \label{ec:SolutionOne}
\mathbf{\Psi}(\mathbf{r},t)
=\exp{\left(i\mathbf{k}\cdot \mathbf{r} \right)}\mathbf{\Phi}(t),
\end{equation}
where $\mathbf{\Phi}(t)= \left(\Phi_{A}(t),\Phi_{B}(t)\right)^{\top}$.
Substituting (\ref{ec:SolutionOne}) reduces
Eq. (\ref{ec:DiracEquationOne}) into
\begin{equation} \label{ec:DiracEquationTwo}
\mathbbm{H}(t)\mathbf{\Phi}(t)=i\hbar\frac{d}{dt}\mathbf{\Phi}(t),
\end{equation}
where the matrix $\mathbbm{H}(t)$ is defined in
Appendix \ref{AppendixA}.

The diagonal terms of $\mathbbm{H}(t)$ can be lifted
by explicitly adding a time-dependent phase to the wave function
\begin{equation}\label{ec:SolutionTwo}
\mathbf{\Phi}(t)=\exp{\left[-\frac{i}{\hbar}\int^t ds
  \alpha_{\mathbf{k}}(s)\right]}{\bm{\chi}}(t), 
\end{equation}
with $\alpha_{\mathbf{k}}(t)=\hbar v_t k_y-e v_t A_y(t)$
and $\bm{\chi}(t)=(\chi_{A}(t),\chi_B (t))^{\top}$.
Following the procedure shown in the Appendix \ref{AppendixA},
Eq. (\ref{ec:DiracEquationTwo}) can be recast in the form
of a second-order ordinary differential equation as
\begin{equation}\label{ec:DiracEquationThree}
\frac{d^2}{dt^2}\bm{\chi}(t)+\mathbbm{F}(t)\bm{\chi}(t)=0,    
\end{equation}
where the function $\mathbbm{F}(t)$ is defined by
\begin{equation}\label{ec:MatrixF}
\mathbbm{F}(t)=-\frac{i}{\hbar}
\hat{\bm{\sigma}}\cdot\frac{d\bm{S}}{dt}
+\frac{1}{\hbar^2}\left[\widetilde{\mathbbm{H}}(t)\right]^2,
\end{equation}
with $[\widetilde{\mathbbm{H}}(t)]^2
=|\boldsymbol{\kappa}_{\boldsymbol{v}}-\boldsymbol{S}|^2$,
$\boldsymbol{\kappa}_{\boldsymbol{v}}=\hbar(v_x k_x, v_y k_y)$
and $\boldsymbol{S}=e(v_x A_x, v_y A_y)$.
In the last expression, the vector $\boldsymbol{\kappa}_{\boldsymbol{v}}$
is the directional energy flux of the electrons, and the components
of $\boldsymbol{S}$ represent the work done by the electromagnetic wave along
the $x$ and $y$ directions.

\section{Elliptically polarized waves}
\label{sec:BoropheEllipticallyPolarized}

Let us now study the case of an elliptically polarized electromagnetic wave
characterized by the vector potential
\begin{equation}\label{ec:VectorPotencial}
{\bf{A}}= \frac{1}{\Omega}\left(E_x\cos(\Omega t),E_y\sin(\Omega t)\right),   
\end{equation}
where $E_x$ and $E_y$ are constants and
$\Omega$ is the frequency of the electromagnetic wave.
The vector potential (\ref{ec:VectorPotencial})
corresponds to the electric field
$\bm{E}=-\partial \boldsymbol{A}/\partial t
=(E_x \sin(\Omega t),-E_y \cos(\Omega t))$.

Rewriting Eq. (\ref{ec:DiracEquationThree})
in terms of the phase
\begin{equation}\label{ec:Phase}
\phi=\Omega t,    
\end{equation}
yields the Hill equation \cite{magnus2013hill}
\begin{equation}\label{ec:HillEquation}
\bm{\chi}''(\phi)+\mathbbm{F}(\phi)\bm{\chi}(\phi)=0, \end{equation}
where $\mathbbm{F}(\phi)$ is
\begin{multline}\label{ec:MatrixFBorophene}
\mathbbm{F}(\phi)=i\left(\frac{\zeta_x}{\hbar\Omega}
\hat{\sigma}_x\sin\phi-\frac{\zeta_y}{\hbar\Omega}
\hat{\sigma}_y\cos\phi\right)\\
+\left(\frac{1}{\hbar\Omega}\right)^{2}
\left[\epsilon^2-2\bm{\kappa}_{\bm{v}}\cdot\bm{S}
+\frac{1}{2}\left(\zeta^2_{x}+\zeta^2_{y}\right)\right.\\
\left.+\frac{1}{2}\left(\zeta^2_{x}-\zeta^2_{y}\right)\cos(2\phi)\right],
\end{multline}
and
\begin{eqnarray}
\bm{S}&=&(\zeta_{x} \cos\phi,\zeta_{x}\sin\phi)\label{ec:vec3},\\
\zeta_{x}&=&eE_{x}v_{x}/\Omega\label{ec:ZetaX},\\
\zeta_{y}&=&eE_{y}v_{y}/\Omega\label{ec:Zetay}.
\end{eqnarray}
The unitless parameter $\epsilon/\hbar\Omega$
is the ratio of the electron energy to the photon energy.
Similarly, the parameter $\zeta_{x}/\hbar\Omega$ 
($\zeta_{y}/\hbar\Omega$) is the ratio of the work done by the
electromagnetic wave along the $x$ ($y$) direction
to the photon energy.   

The determination of the stability regions of
the differential Eq. (\ref{ec:HillEquation})
is quite challenging mainly due to the
imaginary part in the first term of the
right-hand side of Eq. (\ref{ec:MatrixFBorophene}).
{\color{red}While the real part gives rise to the Whittaker-Hill equation}\cite{urwin1970iii}, the imaginary term yields a Mathieu-like equation
with complex characteristic values,
rarely discussed in literature \cite{ZIENER20124513}.
%While the real part gives rise to the standard
%Mathieu differential equation whose stability regions
%are very well known, the imaginary term
%yields a Mathieu-like equation
%with complex characteristic values,
%rarely discussed in literature \cite{ZIENER20124513}.
Fortunately, in the intense electric field
or long wavelength regimes the imaginary part is negligible.
Other limits are treatable by
perturbation theory \cite{lopez2010graphene,higuchi2017light}.

\begin{figure}[t]
\begin{flushleft}
\includegraphics[scale=0.61]{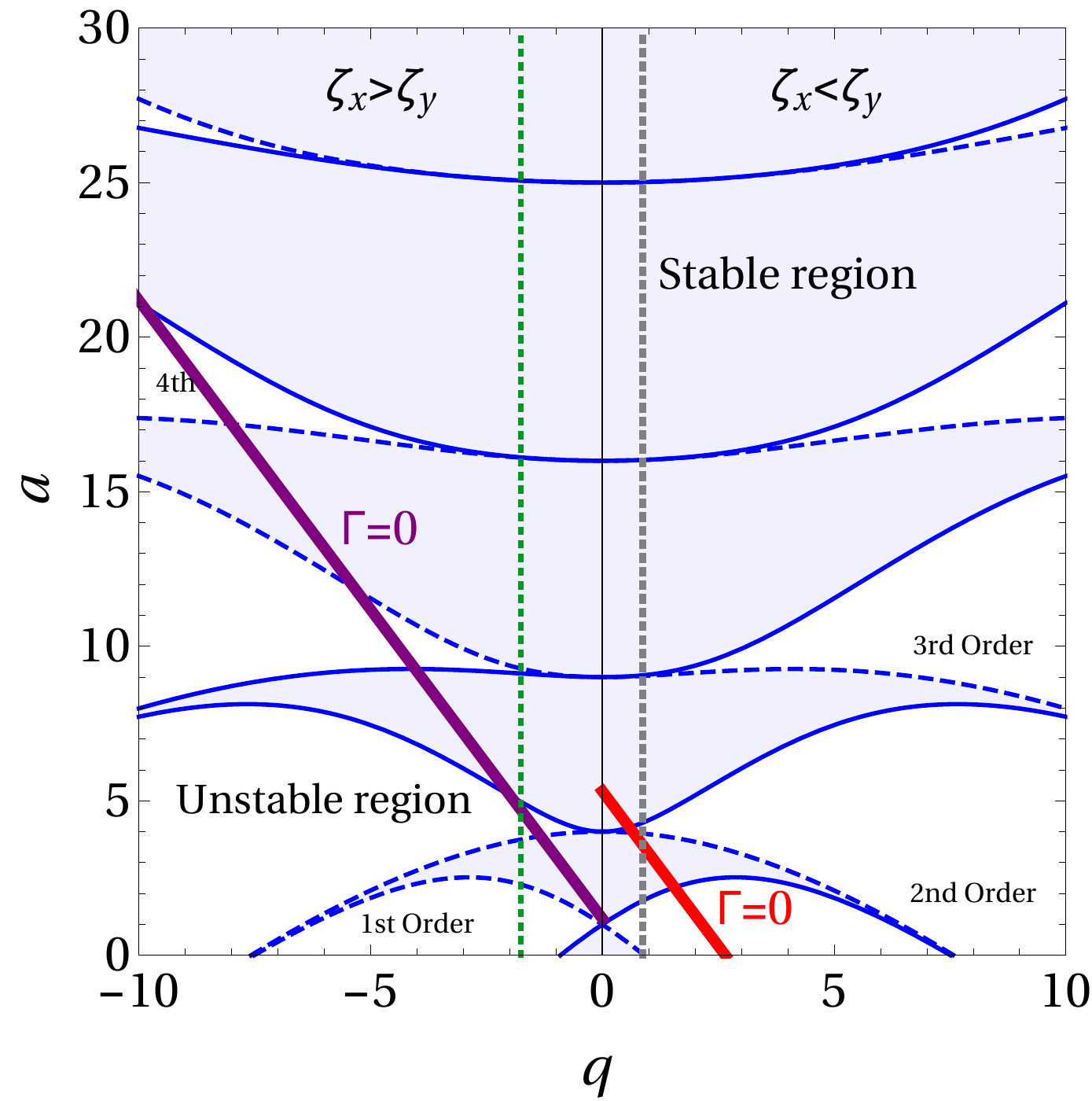}
\end{flushleft} 
\caption{\label{Fig:StabilityMathieu} Mathieu equation stability chart of
in the $q-a$ plane.
The stability (light blue) and instability (white) domains are divided by
the characteristic curves $a_r(q)$ (solid blue lines)
and $b_{r}(q)$ (dashed blue lines) where $r\in \mathbbm{Z}$.
The Mathieu characteristic values $a_r(q)$ and $b_r(q)$ have even parity
with respect to $q$ and therefore
the spectrum is symmetric for $\zeta_x > \zeta_y$
and $\zeta_x < \zeta_y $.
The solid purple ($\zeta_x > \zeta_y$) and
red ($\zeta_x < \zeta_y$) lines
correspond to the extra constraint due to Eq. (\ref{ec:Coef-a}).
The vertical 
dotted green line corresponds to $q=-1.77$
and the dotted gray line to $q=0.87$. }
\end{figure}

Here we focus on the intense electric field regime.
We thus assume that $\zeta_{i}/\hbar\Omega\gg1$ with $i=x,\,y$,
which is equivalent to $ecE_{x}/\hbar\Omega^{2}\gg 349$ and
$ecE_{y}/\hbar\Omega^{2}\gg 435$. This corresponds to
electric fields $E_x\gg1.91\,V/m$ and $E_y\gg2.39\,V/m$.
Thereby, we can neglect the linear terms of $\zeta_{i}/\hbar\Omega$
in Eq. (\ref{ec:HillEquation}) that yield the
imaginary terms.
The obtained expression,
best-known for describing the dynamics of the
parametric pendulum\cite{aldrovandi1980quantum,baker2005pendulum},
is the Mathieu differential equation 
\begin{equation}\label{ec:MathieuEquation}
\bm{\chi}''(\phi)+[a - 2q \cos(2 \phi)]\bm{\chi}(\phi)=0.   
\end{equation}
%as we pointed out before.
The purely real parameters $q$ and $a$ are given by
\begin{eqnarray}
q &=&\frac{\zeta^2_{y}-\zeta^2_{x}}{\left(2\hbar\Omega\right)^2}
  =\left(\frac{e}{2\hbar\Omega^2}\right)^2
  \left(v^2_yE^2_{y}-v^2_xE^2_{x}\right)
  ,\label{ec:Coef-q}\\
a&=&\frac{\epsilon^ 2+\zeta^2_y}{\left(\hbar\Omega\right)^2}-2q
  =\frac{2\epsilon^ 2+\zeta^2_x+\zeta^2_y}{2\left(\hbar\Omega\right)^2}
  \nonumber\\
  &&\,\,\,\,\,= \frac{\epsilon^2}{\left(\hbar\Omega\right)^2}+
     \left(\frac{e}{\hbar\Omega^2}\right)^2
      \left(v^2_xE^2_{x}+v^2_yE^2_{y}\right).\label{ec:Coef-a}
\end{eqnarray}
The characteristic value of the Mehtieu equation
\begin{equation}
\sqrt{a}=\frac{\Omega_0}{\Omega},    
\end{equation}
with $\Omega_0=\sqrt{2\epsilon^2+\zeta^2_x+\zeta^2_y}/\sqrt{2}\hbar$
is the ratio between the fundamental frequency $\Omega_0$ and the 
frequency of the electromagnetic wave $\Omega$.
The characteristic value $a$ parametrizes
the family of ellipses in the $k_x-k_y$ plane
that are characterized by the 
eccentricity $[1-(v_y^2/v_x^2)]^{1/2}$.
Stated differently, each value of the parameter $a$
corresponds to a particular elliptical section of the Dirac cone.
However,
not all the ordered pairs in the $q-a$ plane produce stable solutions
of the Mathieu equation.
Consequently, in the presence of an intense electromagnetic
radiation not all the elliptical sections of the Dirac cones
correspond to stable solutions. In fact, the interaction with
light induces elliptical sections of the
Dirac cone that alternate between forbidden (unstable)
and allowed (stable) solutions.
As can be seen in Fig. \ref{Fig:StabilityMathieu},
the stability chart in the $q-a$ plane
consists of tongue-like stable regions (light blue)
that neighbor  with unstable
regions (white). The Mathieu equation might have either even or
odd stable solutions.
Even stable solutions of (\ref{ec:MathieuEquation}) have the form
\begin{equation}
    \boldsymbol{\chi}(\phi)=\mathcal C(a,q,\phi)
 =\exp[-ir(a,q)\phi]\boldsymbol{f}_C(\phi),\label{eq:floquetsolc}
\end{equation}
where $\mathcal C(a,q,\phi)$ is the even Mathieu function,
$\boldsymbol{f}_C(\phi)$ is an even function with period $\pi$
and $a=a_r(q)$ is the Mathieu even characteristic value.
Conversely, odd stable solutions have the form
\begin{equation}
    \boldsymbol{\chi}(\phi)=\mathcal S(a,q,\phi)
 =\exp[-ir(b,q)\phi]\boldsymbol{f}_S(\phi),\label{eq:floquetsols}
\end{equation}
where $\mathcal S(a,q,\phi)$ is the odd Mathieu function,
$\boldsymbol{f}_S(\phi)$ is an odd function with period $\pi$
and $b=b_r(q)$ is the Mathieu odd characteristic value.
When $r$ is a non-integer rational number, inside the stable regions,
the even and odd characteristic values are identical, namely 
$a_r(q)=b_r(q)$.
The rational function $r(a,q)$ depends on the
Mathieu characteristic value $a$ and the parameter $q$.
On the boundaries between the stable and unstable regions
(solid and dashed blue lines Fig. \ref{Fig:StabilityMathieu})
$r$ takes an integer value and inside the stable regions
$r$ is a non-integer rational number.
Thus, inside the stability regions the even and odd
Mathieu functions have the same characteristic value.
For the particular situation in which $q=0$, (\ref{ec:MathieuEquation})
reduces to the differential equation
of an harmonic oscillator whose solutions are  
$\cos(\sqrt{a}\phi)$ and $\sin(\sqrt{a}\phi)$ \cite{mclachlan1951theory}.
Evidently, in this case $r=\sqrt{a}$.
Moreover, in the special case where $r=\sqrt{a}\in\mathbbm{Z}$,
resonant states are generated for which
\begin{equation}
\frac{\Omega_0}{\Omega}=1,\,2,\,3,...,
\end{equation}
and therefore when $q \to 0$ two contiguous
stability zones are connected.

\begin{figure}[t]
\includegraphics[scale=0.322]{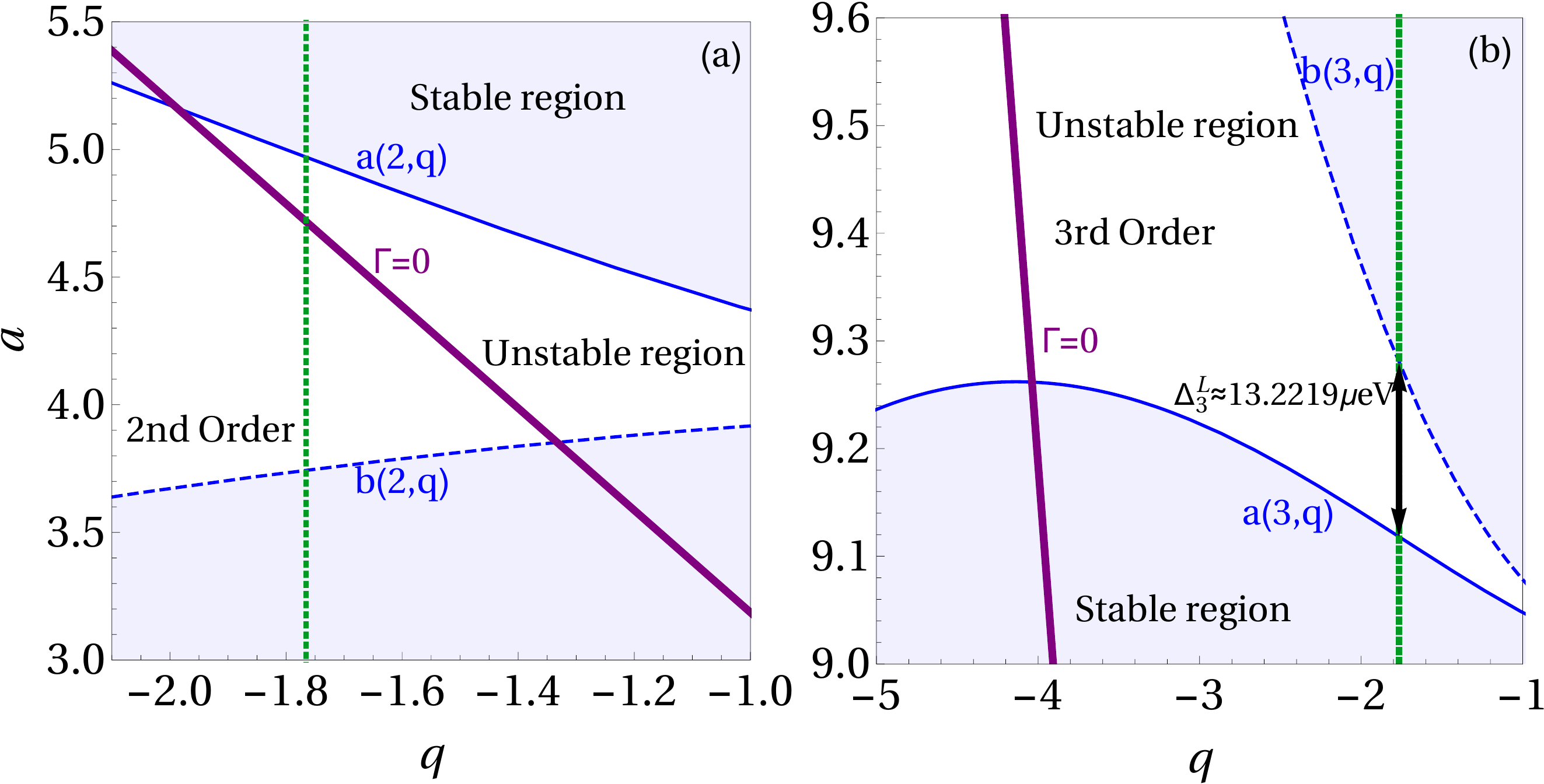}
\includegraphics[scale=0.322]{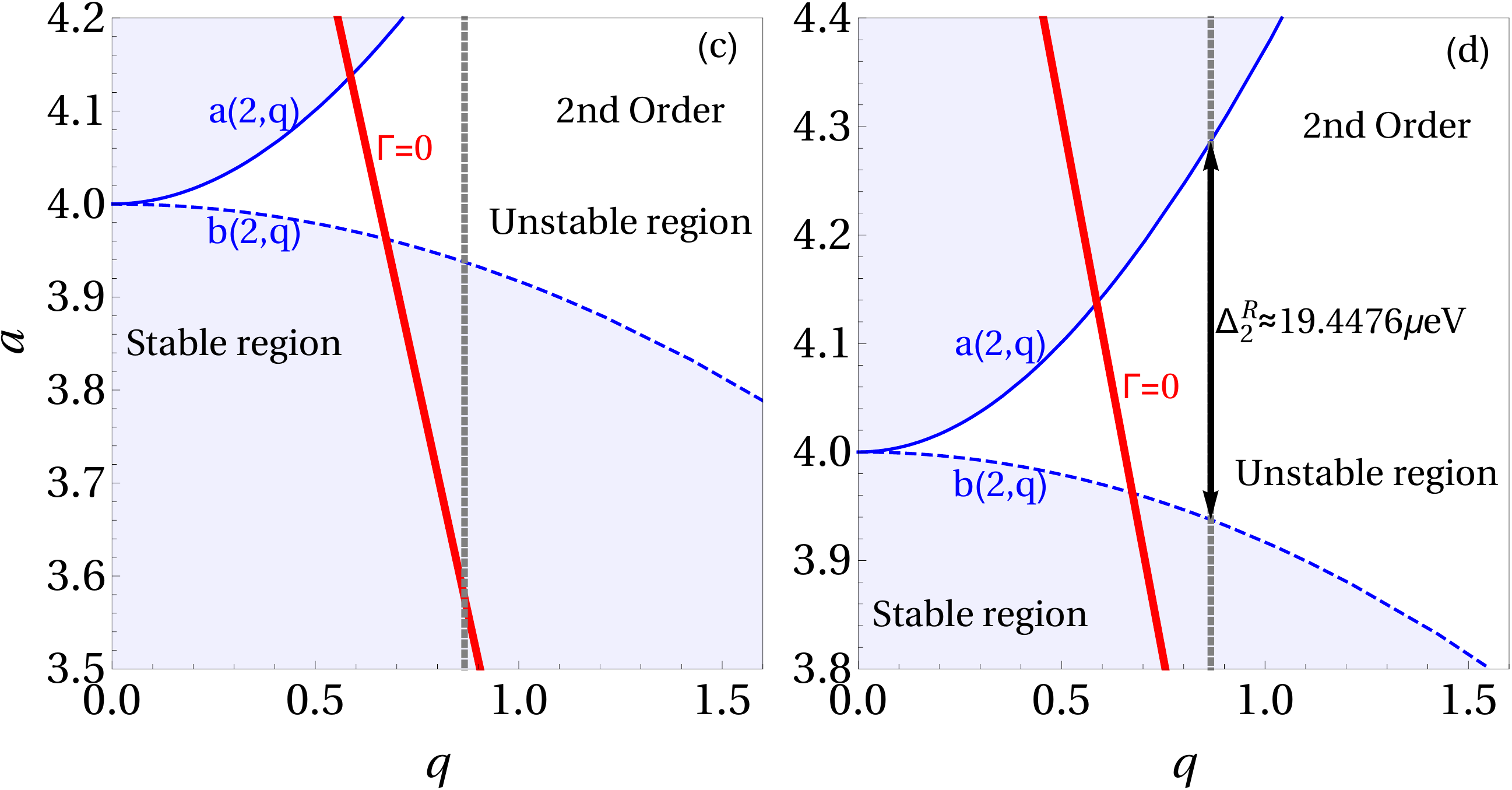}
\caption{\label{Fig:Gap} Zoom of the stability chart near $\Gamma=0$
($\epsilon=0$ and $E^{+}_{\pm 1,k}-\hbar v_t k_y=0$)
in the long wavelength regime.
(a) Crossing of the vertical line $q=-1.77$ (green dotted line)
and condition (\ref{ec:constrainta}) (purple solid line).
The crossing falls inside an unstable state region.
(b) Third-order energy gap $\Delta^{L}_3$ in the
$q<0$ region ($\zeta_x>\zeta_y$)
for fixed fields $E_{x}=5.5$ V/m and $E_y=2.6$ V/m.
(c) Crossing of the vertical line $q=0.87$ (gray dotted line)
and condition (\ref{ec:constrainta}) (red solid line). The
crossing falls inside a stable state region.
(d) Second-order energy gap $\Delta^{R}_2$
in the $q>0$ region ($\zeta_x<\zeta_y$) for fixed fields
$E_{x}=2.6$ V/m and $E_y=5.5$ V/m.
In all the panels the microwave frequency is $\Omega=50$ GHz.}
\end{figure}

\subsection{Wave function and stability spectrum}

The general solution of Eq. (\ref{ec:MathieuEquation}) is
the superposition of the even and odd Mathieu functions 
$\mathcal{C}(a,q,\phi)$ and $\mathcal{S}(a,q,\phi)$.
The wave function is then given by
\begin{multline}\label{ec:WaveFunctiontime}
\boldsymbol{\Psi}(\bm{r},t)
  = \mathcal{N}
  \exp\left[
    i\left(\boldsymbol{k}\cdot \boldsymbol{r}
     -v_t k_y t 
     - \frac{v_t\zeta_y}{v_y\hbar\Omega}\cos(\Omega t)
     \right)\right] \\ 
    \times
    \left[\,\mathcal{C}(a,q,\Omega t)
    \pm i\eta\mathcal{S}(a,q,\Omega t)\,\right]
    \left(\begin{array}{lcc}
         1 \\
        \eta \exp(i\Theta)
    \end{array} \right),
\end{multline}
where $\mathcal{N}$ is a normalization constant,
$\Theta= \tan^{-1}\left(v_{y} k_{y}/v_{x} k_{x}\right)$,
and $\eta=\pm 1$ denotes the conduction and valence bands, respectively.
The wave function (\ref{ec:WaveFunctiontime}) reduces
to the free-particle wave function (\ref{ec:WaveFunctionFree})
when the electric field vanishes.

Since the time-dependent wave function is expressed
in terms of the Mathieu functions, its stability
is governed by the stability chart in \ref{Fig:StabilityMathieu}
that we discussed previously.
Indeed, the structure of the dynamical gaps of Weyl electrons,
generated in the presence of an intense electromagnetic radiation,
is inherited from the properties of the characteristic
values of the Mathieu functions.

The chart can be divided into two key regions according to the
shape of the electromagnetic wave: $q<0$ ($\zeta_x>\zeta_y$) and
$q>0$ ($\zeta_y>\zeta_x$) and $q=0$ ($\zeta_x =\zeta_x$).
If $\zeta_x>\zeta_y$ ($E_x v_x>E_y v_y$) the work done by the
electromagnetic
wave on the electrons is higher along the $x$ axis. Conversely, if  
$\zeta_x<\zeta_y$ ($E_x v_x<E_y v_y$) the work is higher along $y$.
Finally, if $\zeta_x=\zeta_y$ ($E_x v_x=E_y v_y$) the electromagnetic
wave contributes with equal amounts of work in each direction.
Nevertheless, the electron state can not access any point $(q,a)$
in the stability chart shown in Fig. \ref{Fig:StabilityMathieu};
Eq. (\ref{ec:Coef-a}) imposes an extra constraint.
Defining 
$\Gamma=\epsilon/\hbar\Omega=\sqrt{v_x^2k_x^2+v_y^2k_y^2}/\Omega$,
Eq. (\ref{ec:Coef-a}) takes the form
of a straight line $a=\Gamma^2+(\zeta_y/\hbar\Omega)^2-2q$ in the $q-a$
plane. 
Hence, for any state to be accessible to the electron,
the ordered pair $(q,a)$ must
satisfy the inequality
\begin{equation}
a \ge \left(\frac{\zeta_y}{\hbar\Omega}\right)^2-2q.
\label{ec:constrainta}
\end{equation}
In Fig. \ref{Fig:StabilityMathieu} the solid purple and
solid red lines illustrate the limiting case
\begin{equation}
a=(\zeta_y/\hbar\Omega)^2-2q,\label{eq:limitingcase}
\end{equation}
for $\zeta_x>\zeta_y$ and $\zeta_y>\zeta_x$ respectively. 
Naturally, any of these points should also fall on the stable regions
allowed by the Mathieu equation in order to produce a stable
solution of the wave function.

For fixed $E_x$, $E_y$ and $\Omega$, $q$ is constant,
and therefore the allowed states should be located on
the vertical line $q=\mathrm{const}$ (see for example
the green dotted line or the gray dotted line
in Figs. \ref{Fig:StabilityMathieu} and \ref{Fig:Gap}).
Along this lines, the ranges of stable and unstable states alternate
producing the appearance
of bands separated by dynamical energy gaps.
The opening of these gaps is due to the space-time diffraction
of electrons in phase with the electromagnetic field,
and effect akin to the magnetoacoustic diffraction of electrons
in phase with acoustic waves \cite{landau1946vibrations,davydov1980theorie,Champo2019}.

\begin{figure}[t]
\begin{center}
\includegraphics[scale=0.35]{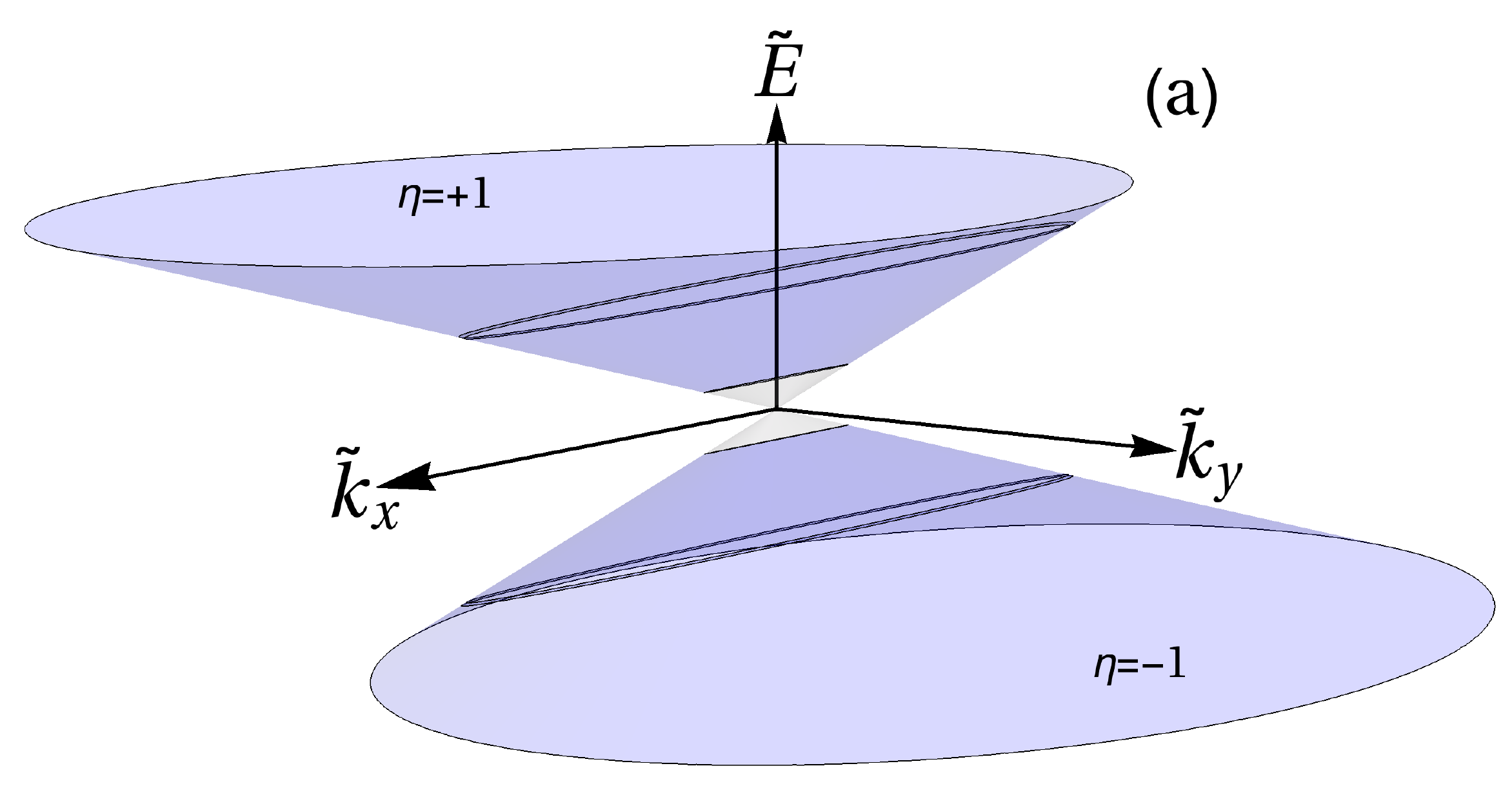}
\includegraphics[scale=0.35]{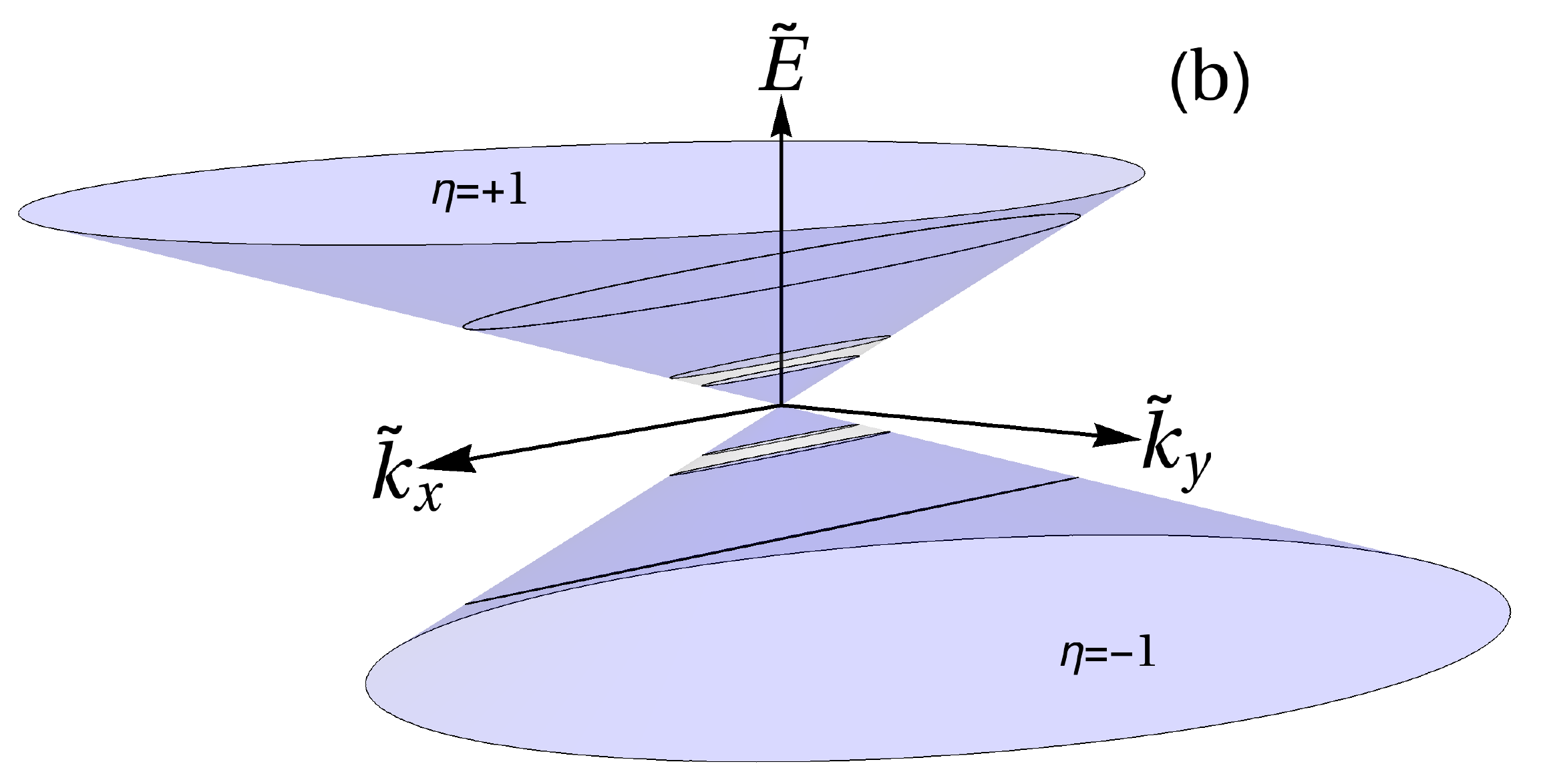}
\end{center} 
\caption{\label{Fig:TiltedDiracCones} Dirac cones-energy dispersion
$\widetilde{E}$
($\widetilde{k}_x,\,\widetilde{k}_y$) ($E^{+}_{\pm 1}/\hbar\Omega$),
for the conduction $\eta=+1$ and valence $\eta=-1$ bands.
(a) A gap opens up a the tip of the Dirac cone
for $\zeta_x>\zeta_y$ ($E_x=5.5$ V/m and $E_y=2.6$ V/m).
(b) When $\zeta_x<\zeta_y$ ($E_x=2.6$ V/m and $E_y=5.5$ V/m) gaps
only open up far from Dirac point.
The white regions correspond to forbidden energies and the blue ones
to the allowed energies.}
\end{figure}

To further comprehend the connection
between the Mathieu stability chart
and the consequent wave function gap structure,
it is illustrating to project
the stable and unstable regions
of Fig. \ref{Fig:StabilityMathieu}
on the surface of the tilted Dirac cones
that arise from the free particle Weyl equation.
To this end, we explicitly express the normalized energy dispersion
$\widetilde{E}=(1/\hbar\Omega)E^{+}_{\pm 1, k}$ from
Eq. (\ref{ec:EnergyDispersion})
in terms of the normalized wave vector components 
$(\widetilde{k}_{x},\widetilde{k}_{y})$
and the parameter $a$ in the Eq. (\ref{ec:Coef-a})
obtaining
\begin{equation}\label{ec:ElipticRings}
\left(\frac{v_x}{v_F}\right)^{2}\widetilde{k}^2_{x}
+\left(\frac{v_y}{v_F}\right)^{2}\widetilde{k}^{2}_{y}
=a+2q-\left(\frac{\zeta_y}{\hbar\Omega}\right)^2, 
\end{equation}
where $\widetilde{k}_{x}= (v_F/\hbar\Omega) k_x$ and
$\widetilde{k}_{y}= (v_F/\hbar\Omega) k_y$.
Elliptical rings of allowed and forbidden states form in the
$(\widetilde{k}_{x},\widetilde{k}_{y})$ plane or on the
surface of the Dirac cone
for fixed values of $E_x$ and $E_y$
(or fixed values of $q$ and $\zeta_y/\hbar\Omega$).
The dressed Dirac cones, the Dirac cones over whose
surfaces the allowed and forbidden states have been projected,
are shown in Fig. \ref{Fig:TiltedDiracCones}.
The light blue portion of the surface represents the allowed states
and the white rings are the forbidden ones. The first
correspond to the stable regions and the latter to the unstable
regions
of Fig. \ref{Fig:StabilityMathieu}.

In Figs. \ref{Fig:Gap} (a) and (b) we plot
the vertical line $q= - 1.77$ (green dotted line) and
the line (\ref{eq:limitingcase})
(solid purple line) superimposed to a zoom
of the stability chart
for typical electromagnetic
field values $E_x=5.5$ V/m, $E_y=2.6$ V/m and
$\Omega=50$ GHz.
The crossing between these two lines,
seen in Fig. \ref{Fig:Gap} (a), is the starting point
for the search of stable solutions.
However, in the immediate region above the crossing we observe
a gap of unstable solutions, that projected on to the
Dirac cone produces the appearance of forbidden
states at the tip, forming a gap.
At higher energies, we observe the rings corresponding to
the third order gap as can be appreciated in
Figs. \ref{Fig:StabilityMathieu} and \ref{Fig:Gap} (b).
This gap yields
and energy range $\Delta^L_3=13.22\,\mu$eV
(see Appendix \ref{AppendixB}) of forbidden states.
It should be noted that the origins of the first gap at the
Dirac point and the following ring-like forbidden regions
are essentially the same.
Both of them are generated in points that comply with
the inequality (\ref{ec:constrainta}),
and as a result of
inherent instabilities of the Mathieu solutions.

When the parameters are chosen to
fall on the opposite side of the stability chart ($\zeta_x<\zeta_y$)
the arrangement of the gaps is quite different.
In Fig. \ref{Fig:Gap} (c) we observe the
crossing of the vertical line $q=0.87$
and the limiting line (\ref{eq:limitingcase})
for electromagnetic field values $E_x=2.6$ V/m, $E_y=5.5$ V/m and
$\Omega=50$ GHz.
In contrast to the previous case, above the crossing
we find ourselves well inside a stability region.
Hence, the tip of the Dirac cone
is dressed entirely with allowed states and the forbidden rings
appear well above it as can be seen
in Fig. \ref{Fig:TiltedDiracCones}.
At high energies the line $q=0.87$ crosses the second order gap
as can be seen in Fig. \ref{Fig:Gap} (d).
A ring of unstable states with an energy gap
of $\Delta^R_2=19.45\,\mu$eV is projected on to the Dirac cone
until the line reaches the next stability zone
[see Fig. \ref{Fig:TiltedDiracCones} (b)].

\begin{figure}[t]
\begin{flushleft}
\includegraphics[scale=0.58]{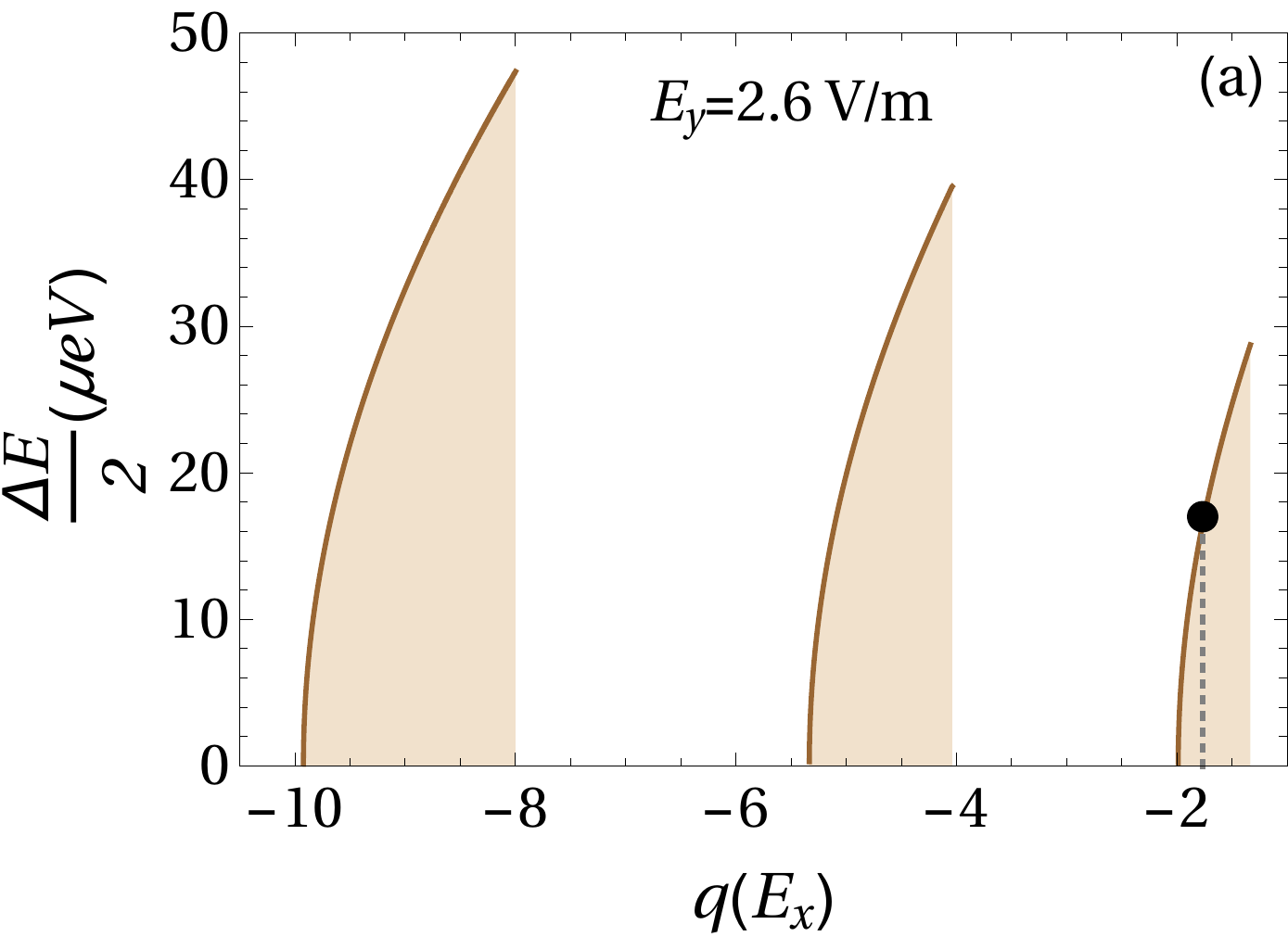}
\includegraphics[scale=0.6]{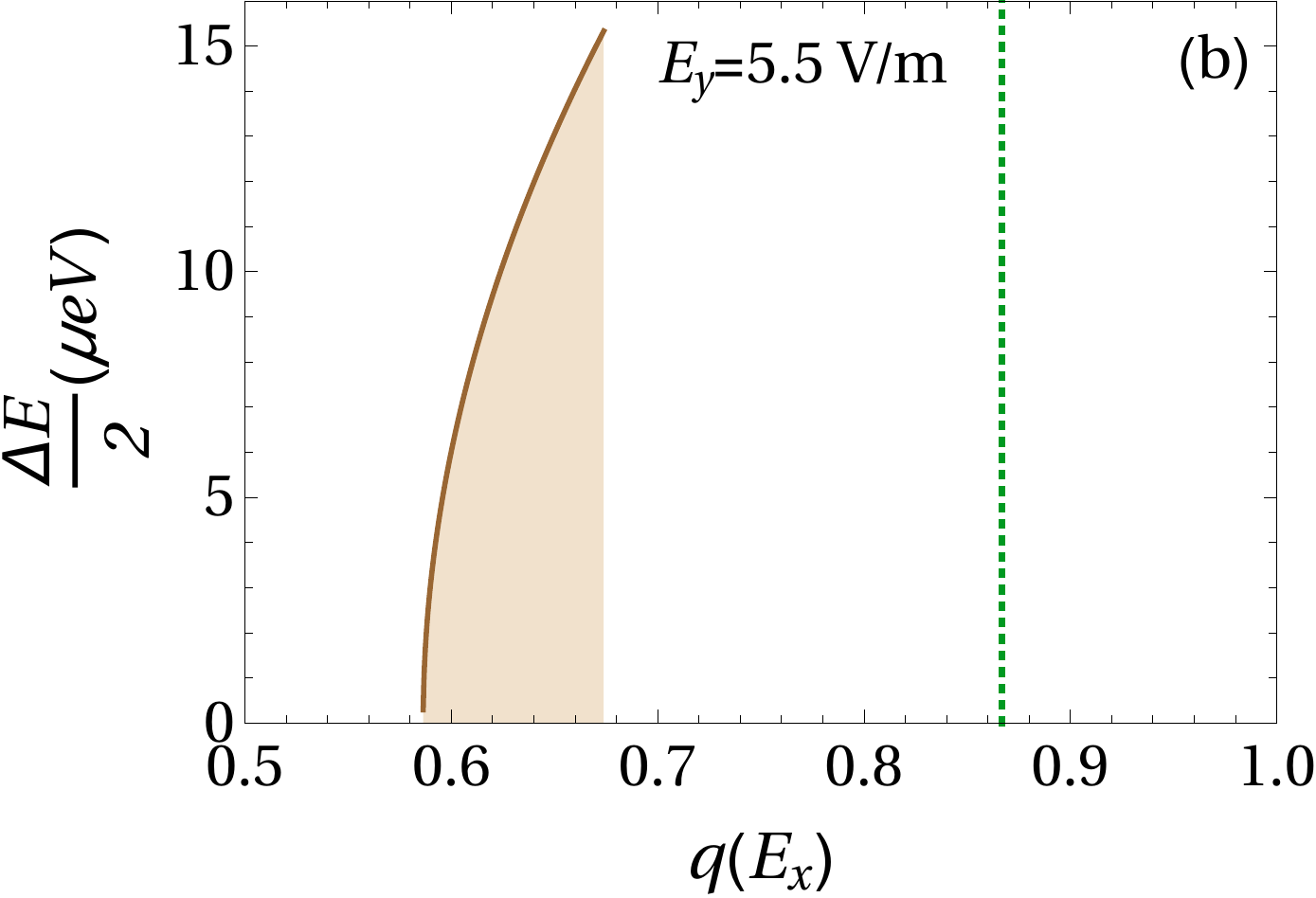}
\end{flushleft}
\caption{\label{Fig:HalfGapEnergy} Energy gap ($\Delta E(q)/2$)
regions near the Dirac points.
The brown solid lines represent $\Delta E/2$ as
a function of $q$ in the domains where $\Delta E(q)>0$
for (a) $E_y=2.6$ V/m and $\zeta_x >\zeta_y$,
and (b) $E_y=5.5$ V/m and $\zeta_x<\zeta_y$.
The vertical dotted gray line and the dot in $q=-1.77$ (a)
correspond to
$E_x=5.5$ V/m, $\Delta E/2=17\, \mu $eV. Likewise,
the vertical dotted green line corresponds to $q=0.87$ (b).}
\end{figure}

To systematize the search of gaps in the $K$ point
of the Dirac cone we define
the indicator
\begin{equation}
\Delta E(q)/2=\hbar\Omega\sqrt{c_r(q)
  -\left[\left(\frac{\zeta_y}{\hbar \Omega}\right)^2-2 q\right]}, 
\end{equation}
where $r\in \mathbbm{Z}$ and $c_r(q)$ is either $a_r(q)$
or $b_r(q)$, depending on which one is at the bottom
of the allowed band.
This indicator corresponds to the energy difference between
the lower allowed band edge and the limiting case of
the inequality (\ref{ec:constrainta}) given by (\ref{eq:limitingcase}).
The integer $r$ is chosen so that the
purple line in Fig. \ref{Fig:StabilityMathieu}
is situated directly below
the top band edge associated with the stable region.
Therefore, if for a given value of $q$
the purple line falls on a forbidden region
of states then $\Delta E(q)>0$. If, on the other hand,
the purple line falls on an allowed band $\Delta E(q)$ is
a pure imaginary number. The domain where $\Delta E(q)$ is a
pure real number corresponds, thus, to a gap of forbidden states.
Hence, the function $E(q)$ provides with a clear-cut
criterion to detect the formation of gaps in
the surroundings of the Dirac point:
$\Delta E(q)\in \mathbbm{R}$.

In Fig. \ref{Fig:HalfGapEnergy} we analyze
the gap formation through the behaviour of $\Delta E(q)$
in the case $\zeta_x>\zeta_y$ for fixed $E_y=2.6$V/m.
The domains of $q$ where $\Delta E(q)>0$ are shown as solid brown lines
in Fig. \ref{Fig:HalfGapEnergy} (a).
The point given by $E_x=5.5$ V/m, $q=-1.77$,
examined previously in Fig. \ref{Fig:TiltedDiracCones} (a),
is shown in Fig. \ref{Fig:HalfGapEnergy} (a).
We notice that this value of $q$ falls inside one of the
domains where $\Delta E(q)>0$,
therefore indicating the presence of a gap opening at the tip
of the cone. Moreover, another useful property of $\Delta E(q)$
is that it gives the energy of the gap.
In this example $\Delta E(-1.77)/2=17\mu$eV
which corresponds to the gap shown in Fig. \ref{Fig:TiltedDiracCones}(a).
In contrast, for $E_x=2.6$ V/m, $E_y=5.5$ ($\zeta_x<\zeta_y$)
and $q=0.89$, Fig. \ref{Fig:HalfGapEnergy} (b) shows
that there is no gap formation as expected from
Fig. \ref{Fig:TiltedDiracCones}.

\begin{figure}
\includegraphics[scale=0.5]{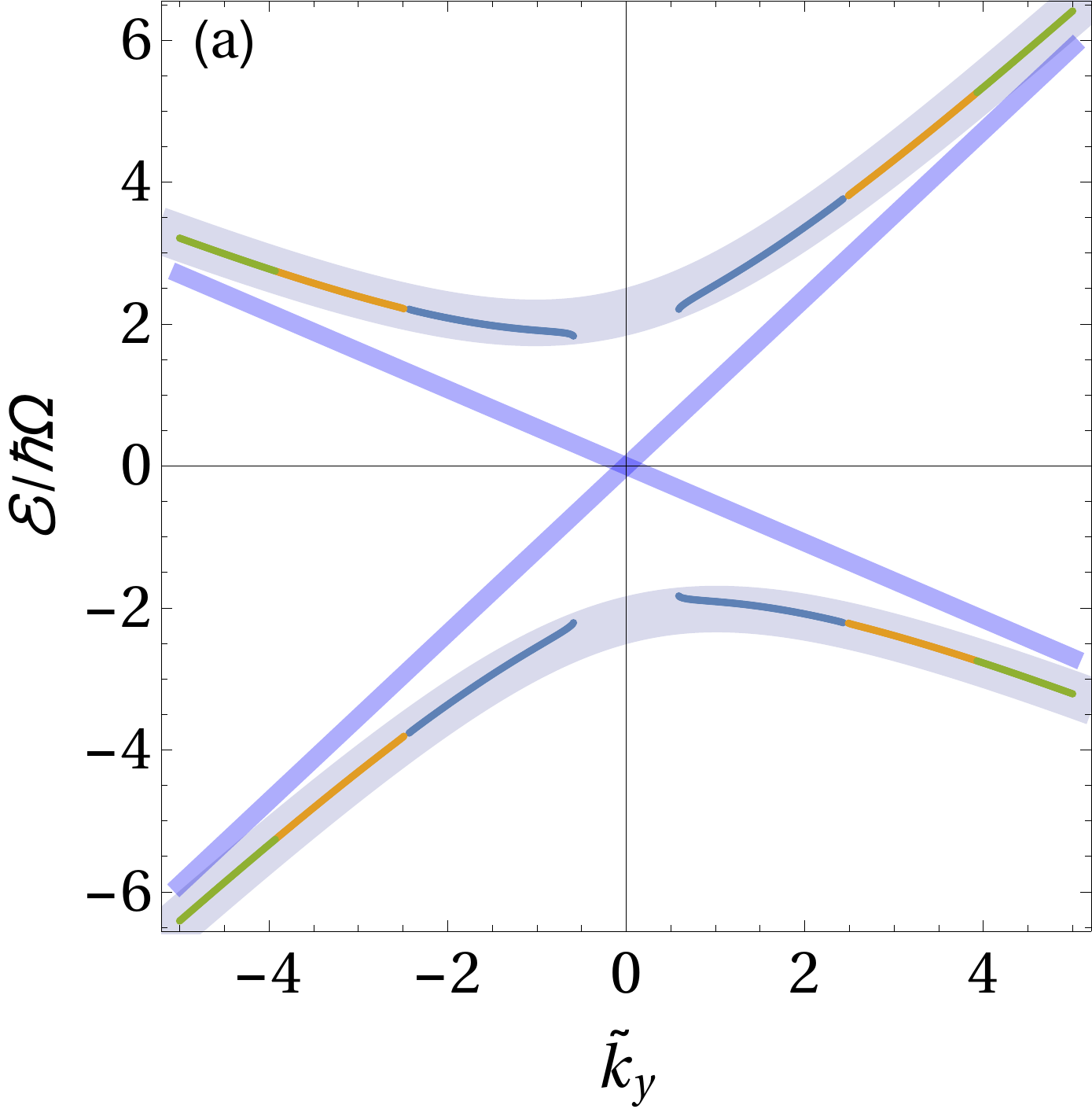}
\includegraphics[scale=0.5]{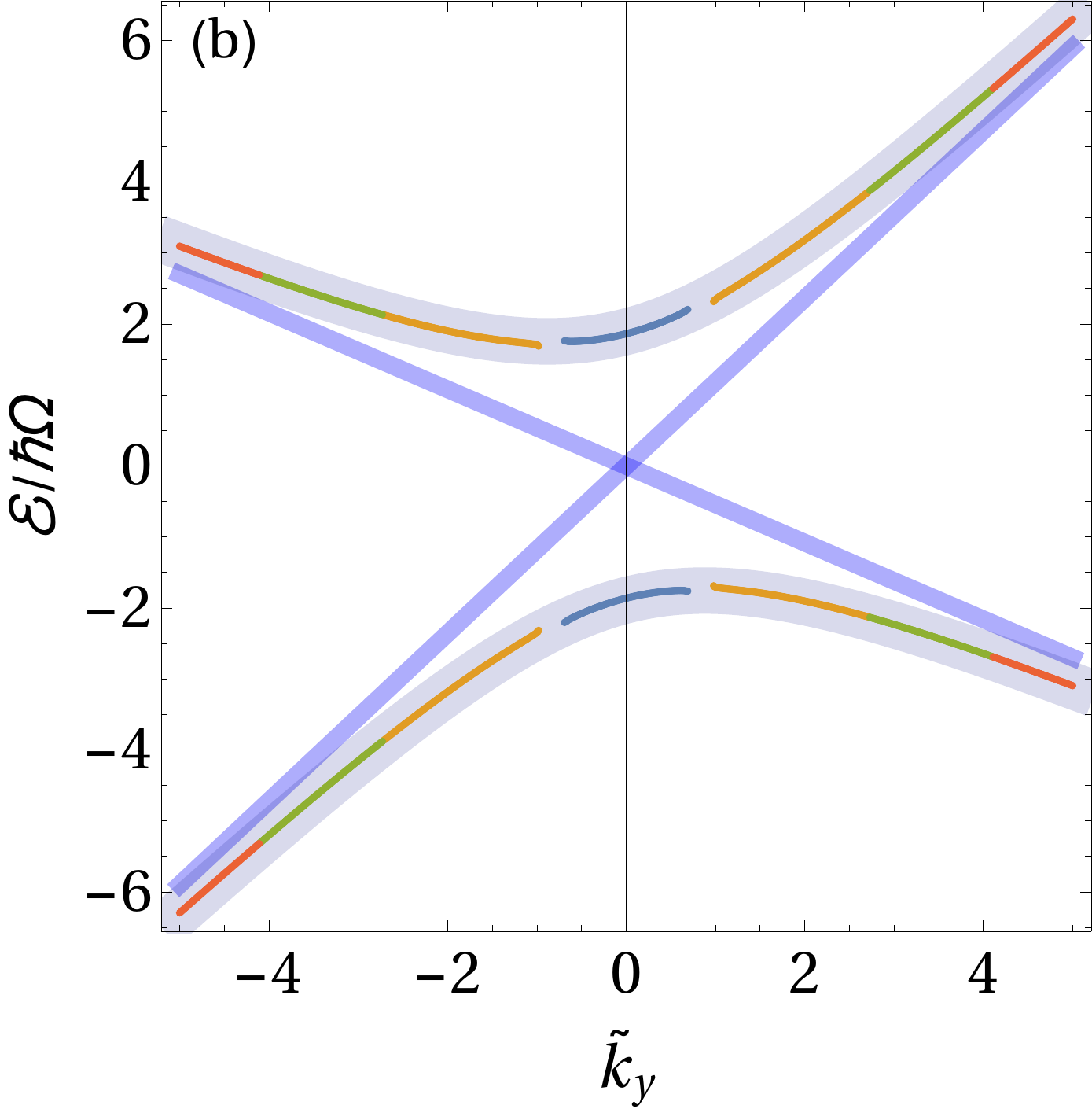}
\caption{\label{Fig:sectionquasienergy} Cut of the quasi-energy
spectrum with the section plane $k_x=0$
as a function of the pseudo-momentum $\widetilde{k}_y$
for (a) $E_x=5.5$ V/m, $E_y=2.6$ V/m, $q=-1.77$ ($\zeta_x >\zeta_y$),
and (b) $E_x=2.6$ V/m, $E_y=5.5$ V/m, $q=0.87$  ($\zeta_x<\zeta_y$).
The bands of allowed states are distinguished with different colors:
blue for $r=1$, orange for $r=2$, green for $r=3$ and
red for $r=4$. The light blue solid lines are sections of the
Dirac cone. The section
of the quasi-energy spectrum in the approximation where $r=\sqrt{a}$
is shown for reference as a solid gray line.
}
\end{figure}

\begin{figure}
\includegraphics[scale=0.35]{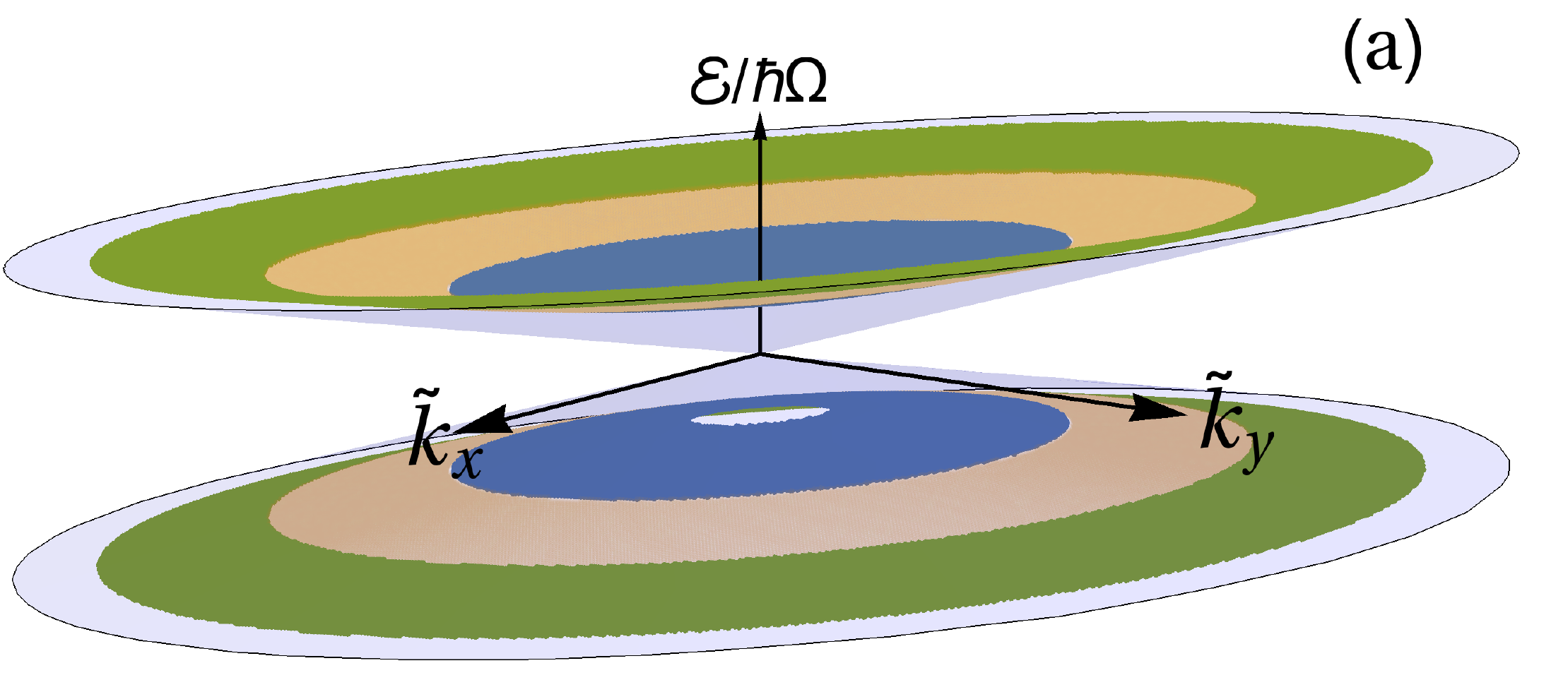}
\includegraphics[scale=0.35]{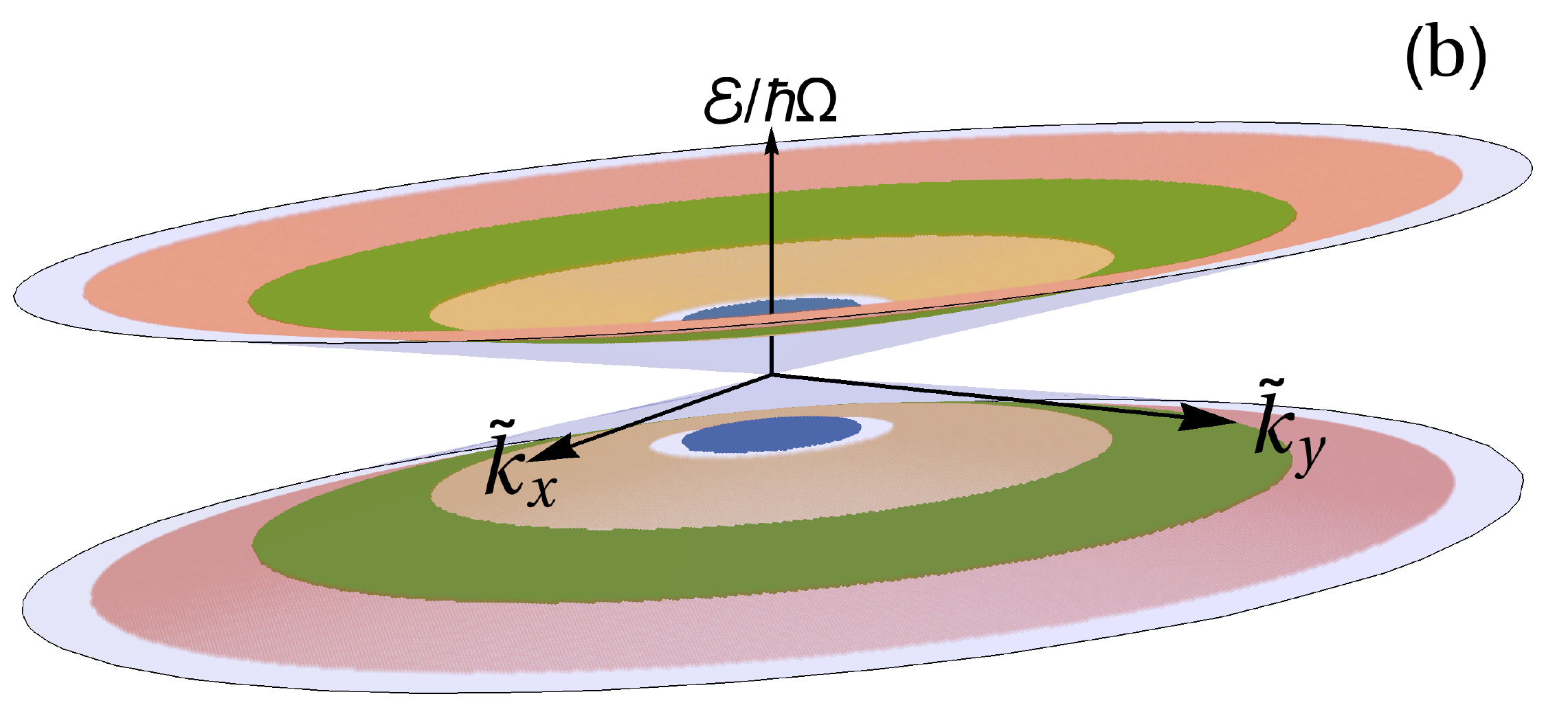}
\caption{\label{Fig:quasienergy} Quasi-energy
spectrum as a function of the pseudo-momentum components
$\widetilde{k}_x$ and $\widetilde{k}_y$
for (a) $E_x=5.5$ V/m, $E_y=2.6$ V/m, $q=-1.77$ ($\zeta_x >\zeta_y$),
and (b) $E_x=2.6$ V/m, $E_y=5.5$ V/m, $q=0.87$  ($\zeta_x<\zeta_y$).
The bands of allowed states are distinguished with different colors:
blue for $r=1$, orange for $r=2$, green for $r=3$ and
red for $r=4$.
}
\end{figure}
These results clearly show that if $E_x>E_y$, the function
$\Delta E(q)/2$ presents more domains that yield forbidden gaps
close to the Dirac points than $E_y>E_x$.
This strongly suggests that
the shape and position of the gaps is strongly influenced
by the relative orientation of the minor and major axes of
the elliptical profiles of the radiation and the
free-Weyl particle Dirac cone. If the major and minor
axes of the ellipses arising from the Dirac cone are perpendicular
to the major and minor axes of the ellipses of
the electromagnetic profile, a gap
at the Dirac point is more likely to form.
Otherwise, if the ellipses are oriented
in the same direction, gaps are more improbable.
It is important to emphasize that this does not
imply that the two ellipses necessarily have to have the
the same proportions.
Nevertheless, the only way to correctly predict the formation
of a gap in the Dirac point is to determine
if the value of $q$ falls inside one of the domains
were $\Delta E>0$, as we discussed above.

\subsection{Quasi-energy spectrum}\label{sec:quasienergy}

The Hamiltonian in Schr\"odinger equation (\ref{ec:DiracEquationOne})
is a periodic function of time,
therefore the Floquet theorem must hold
and, consequently, the wave function must be of the form
\begin{equation}
    \boldsymbol{\Psi}(\boldsymbol{r},t)
    =\mathcal{N}\exp\left(-i\frac{\mathcal{E}t}{\hbar}\right)
    \boldsymbol{f}(\Omega t).\label{eq:quasipsi}
\end{equation}
The phase $\mathcal{E}$ is usually termed the quasi-energy and
$\boldsymbol{f}(\Omega t)=\boldsymbol{f}(\Omega t+2\pi)$
is a periodic function of time with the same period
$T=2\pi/\Omega$ as the Hamiltonian.
Using Eqs. (\ref{eq:floquetsolc}) and (\ref{eq:floquetsols})
we can rearrange
the Mathieu functions as
\begin{equation}
\mathcal{C}(a,q,\Omega t)-i\eta \mathcal{S}(a,q,\Omega t)=
\exp[-i \eta r(a,q) \Omega t]f(\Omega t),\label{eq:phasefloquet}
\end{equation}
where $r(a,q)$ is purely rational and therefore $a_r=b_r$ as we
discussed before.
Substituting Eq. (\ref{eq:phasefloquet}) into
(\ref{ec:WaveFunctiontime}) and
comparing with (\ref{eq:quasipsi})
we find the explicit expression for the
quasi-energy
\begin{equation}\label{eq:exactquasienergy}
\frac{\mathcal{E}}{\hbar \Omega}= \frac{v_tk_y}{\Omega}+\eta r(a,q).
\end{equation}

To better visualize the shape of the spectrum
we may use $r\approx \sqrt{a}$ for $r\gg 1$
to approximate the quasi-energy. Using
the definition of $a$ (\ref{ec:Coef-a}) and taking the limit
for large quasi-momenta we get
\begin{equation}
    \lim_{k_x,k_y\rightarrow \infty}\mathcal{E}=
    \hbar v_tk_y+\hbar\sqrt{v_{x}^{2} k_{x}^{2}+v_{y}^{2}k_{y}^{2}}.
\end{equation}
Hence, the quasi-energy spectrum asymptotically approaches
the Dirac cone for large quasi-momentum values.
This feature is clearly seen
in Fig. \ref{Fig:sectionquasienergy}. The light blue
straight lines are the section of the Dirac cone
cut by the $k_x=0$ plane and the solid gray lines are
the quasi-energy spectrum in the approximation $r=\sqrt{a}$.
The blue, orange, green and red lines correspond to the
different bands arising from the exact expression
of the quasi-energy (\ref{eq:exactquasienergy}).
We readily confirm that
both, the exact and the approximated quasi-energy spectrum lines,
asymptotically come close to the Dirac cone.
This is also seen in Fig. \ref{Fig:quasienergy}, where
the full quasi-energy surface is shown and
the Dirac cone is depicted as a light blue semi-transparent surface.
The most striking characteristic of
these plots is the formation of gaps.
In Fig. \ref{Fig:sectionquasienergy} (a) a gap
appears close to the Dirac point for $E_x=5.5$ V/m, $E_y=2.6$ V/m
and $q=-1.77$.
It is located inside the same region
as the forbidden states in the dressed Dirac cone 
shown in Fig. \ref{Fig:TiltedDiracCones} (a).
In the full spectrum of Fig. \ref{Fig:quasienergy} (a)
this gap translates into a disc-shaped vacuum of states in the tip
of the quasi-energy spectrum.
Instead, for $E_x=2.6$ V/m, $E_y=5.5$ V/m and $q=0.87$
there is no gap formation close to the Dirac point,
though,  a ring-shaped gap appears around the tip of the quasi-energy
spectrum as it is shown in
Figs. \ref{Fig:sectionquasienergy} (b) and \ref{Fig:quasienergy} (b).
These results are consistent with the ones found for the dressed
Dirac cones. Newly, the relative orientation of the minor and
major axes of the radiation and the Dirac cone determine
the location and structure of the gaps.

\section{Conclusions}\label{sec:Conclusions}

We have systematically investigated the wave function stability
of charge carries in a Weyl Hamiltonian under an intense
elliptically polarized electromagnetic radiation.
To this end, we have worked out the time-dependent
wave function from the Schr\"odinger equation in the
limit of strong electric field (or long wavelength).
We have proven that the stability properties of the wave
functions are inherited from the Mathieu functions,
in terms of which they are expressed.
The analysis of the stability chart of the Mathieu functions
projected onto the Dirac cones shows the formation of gaps
of unstable states for certain domain regions of the quasi-momentum
space. We have shown that the structure of the gaps
strongly depends on the alignment between the minor and major axes
of the elliptical profiles of the radiation and the Dirac cone
of the free-Weyl particle spectrum. In summary, the formation
of a gap at the Dirac point is more likely if the radiation
and Dirac cone axes do not match, or are perpendicular.
Otherwise, if the
axes are aligned, ring-shaped gaps form
for higher energies.
The quasi-energy spectrum extracted from the phase of the
wave function consistently reproduces the position and shapes
of these gaps.
Magnitude estimations of the electromagnetic fields
and gap values were presented for the $8-Pmmn$ borophene phase.

\section{Acknowledgements}\label{sec:Acknowledgements}

This work was supported by DCB UAM-A grant numbers
2232214 and 2232215, and UNAM DGAPA PAPIIT IN102717.
 V.G.I.S and J.C.S.S. acknowledge the total support from 
DGAPA-UNAM fellowship. 

\appendix

\section{}\label{AppendixA}
In this Appendix, we derive Eqs. (\ref{ec:DiracEquationOne}) and
(\ref{ec:DiracEquationThree}) from the ansatz (\ref{ec:SolutionOne}) 
and (\ref{ec:SolutionTwo}). We start from  the Dirac equation 
(\ref{ec:DiracEquationOne}) and  applying the solution 
(\ref{ec:SolutionOne}), we obtain Eq. (\ref{ec:DiracEquationTwo}), 
where
\begin{equation}\label{ec:HamiltonianMatrixOne}
\mathbbm{H}(t)= 
\begin{pmatrix}
\alpha_{\mathbf{k}} & \beta^{*}_{\mathbf{k}}\\
\beta_{\mathbf{k}} & \alpha_{\mathbf{k}}
\end{pmatrix},
\end{equation}
whose matrix elements are given by
\begin{eqnarray}
&&\alpha_{\mathbf{k}}=\hbar v_t k_y- v_t e A_y,\label{ec:ExpressionAlpha}\\
&&\beta_{\mathbf{k}}=\hbar(v_x k_x+i v_y k_y)-e[v_x A_x+i v_y A_y].\label{ec:ExpressionBeta}
\end{eqnarray}
In the previous equatins,
the vector components $A_x$ and $A_y$ only depend on
the time variable. 

Now, we substitute Eq. (\ref{ec:SolutionTwo}) in Eq. 
(\ref{ec:DiracEquationTwo}), to get
\begin{equation} \label{ec:DiracEquationThreeApen}
\mathbbm{\widetilde{H}}(t)\bm{\chi}(t)
=i\hbar\frac{d}{dt}\bm{\chi}(t),    
\end{equation}
where
\begin{equation}\label{ec:HamiltonianMatrixTwoApen}
\mathbbm{\widetilde{H}}(t)= 
\begin{pmatrix}
0 & \beta^{*}_{\mathbf{k}}\\
\beta_{\mathbf{k}} & 0
\end{pmatrix}.
\end{equation}
From Eq. (\ref{ec:DiracEquationThreeApen}) we obtain
\begin{equation} \label{ec:DiracEquationFourApen}
\frac{d^2}{dt^2}\bm{\chi}(t)
+\frac{i}{\hbar}\left[\frac{d}{dt}
\mathbbm{\widetilde{H}}(t)\right]\bm{\chi}(t)
+\frac{1}{\hbar^{2}}
\left[\mathbbm{\widetilde{H}}(t)\right]^2\bm{\chi}(t)=0,
\end{equation}
where
\begin{equation}
\frac{d}{dt}\mathbbm{\widetilde{H}}(t)
=-\hat{\bm{\sigma}}\cdot\frac{d\bm{S}}{dt}.
\end{equation}
In the equations above we used the vectors 
$\hat{\bm{\sigma}}=(\hat{\sigma}_x, \hat{\sigma}_y)$ and
$\bm{S}=e(v_x A_x, v_y A_y)$. Therefore, Eq. 
(\ref{ec:DiracEquationFourApen}) can be reduced into
\begin{equation}\label{ec:DiracEquationFiveApen}
\frac{d^2}{dt^2}\bm{\chi}(t)+\mathbbm{F}(t)\bm{\chi}(t)=0,    
\end{equation}
with
\begin{equation}
\mathbbm{F}(t)=-\frac{i}{\hbar}\hat{\bm{\sigma}}
\cdot\frac{d\bm{S}}{dt}+\frac{1}{\hbar^2}
\left[\mathbbm{\widetilde{H}}(t)\right]^2,
\end{equation}
and
\begin{eqnarray}
\left[\widetilde{\mathbbm{H}}(t)\right]^2
=\left|\bm{\kappa}_{\bm{v}}-\bm{S}\right|^2,    
\end{eqnarray}
where $\boldsymbol{\kappa}_{\boldsymbol{v}}=\hbar(v_x k_x, v_y k_y)$.

\section{}\label{AppendixB}
To estimate the gap size, first,
we calculate the value of the parameter $q$ from the Eq. 
(\ref{ec:Coef-q}), and subsequently use the expression 
\cite{lopez2010graphene,Champo2019}
\begin{equation}\label{ec:Gap-n-order}
\Delta_{r}=\hbar\Omega\sqrt{|b_r(q)-a_r(q)|},     
\end{equation}
where $a_{r}(q)$ and $b_{r}(q)$ for $r\in \mathbbm{Z}$
are the boundaries of a forbidden region as it is shown
in Fig. \ref{Fig:StabilityMathieu}.  

For a microwave with frequency $\Omega= 50$ GHz,
we estimate the gap in the following cases:
a) For $E_{x}=5.5$ V/m and $E_{y}=2.6$ V/m ( $\zeta_x>\zeta_y$),
we find $q=-1.77$ and $\Delta^{L}_3 = 13.22\, \mu$ eV. 
b) For $E_{x}=2.6$ V/m and  $E_{y}=5.5$ V/m ($\zeta_x<\zeta_y$),
we find $q=0.87$ and $\Delta^{R}_2 = 19.45\, \mu$ eV. 

%\bibliography{Biblio.bib}

%

\end{document}